# Effects of Spatial Heterogeneity in Rainfall and Vegetation Type on Soil Moisture and Evapotranspiration

Michael J. Puma[*,1], Michael A. Celia[1], Ignacio Rodriguez-Iturbe[1], Jan M. Nordbotten[2,1], Andrew J. Guswa[3], Dmitri Kavetski[1]

[1]*Department of Civil & Environ. Eng., Princeton University, Princeton, New Jersey, 08544, USA*
[2]*Department of Mathematics, University of Bergen, Joh. Bruns gate 12, 5008 Bergen, Norway*
[3]*Picker Engineering Program, Smith College, 51 College Lane, Northampton, MA 01060*

---

**Abstract:**

Nonlinear plant-scale interactions controlling the soil-water balance are generally not valid at larger spatial scales due to spatial heterogeneity in rainfall and vegetation type. The relationships between spatially averaged variables are hysteretic even when unique relationships are imposed at the plant scale. The characteristics of these hysteretic relationships depend on the size of the averaging area and the spatial properties of the soil, vegetation, and rainfall. We upscale the plant-scale relationships to the scale of a regional land-surface model based on simulation data obtained through explicit representation of spatial heterogeneity in rainfall and vegetation type. The proposed upscaled function improves predictions of spatially averaged soil moisture and evapotranspiration relative to the effective-parameter approach for a water-limited Texas shrubland. The degree of improvement is a function of the scales of heterogeneity and the size of the averaging area. We also find that single-valued functions fail to predict spatially averaged leakage accurately. Furthermore, the spatial heterogeneity results in scale-dependent hysteretic relationships for the statistical-dynamic and Montaldo & Albertson approaches.

* Corresponding author (now at Columbia University). Tel: 1-212-678-5667
Current e-mail address: mjp38@columbia.edu

## 1. Introduction

The dynamic linkage among soil, plants, and atmosphere through the temporal and spatial variability of soil moisture is a fundamental component of atmospheric, hydrological, and ecological models [23, 44]. The relationships controlling soil-moisture dynamics are strongly dependent on spatial and temporal scales [44]. Accordingly, a critical issue for efforts to improve model predictions of land-surface processes is the disparity between the resolution of land-surface models and the spatial scale at which physical relationships controlling soil-moisture dynamics are known [10]. Regional land-surface models have resolutions of $\sim 10^2$ km$^2$, while general circulation models (GCMs) are typically run at resolutions of $\sim 10^4$ km$^2$ [16]. In addition, the spatial scale of forcing data for land-surface models from next-generation spaceborne passive microwave sensors will be $\sim 10^3$ km$^2$ at best [10, 25]. Heterogeneity in soil properties, vegetation, topography, rainfall, and atmospheric conditions can be pronounced at these large spatial scales.

The effects of spatial heterogeneity in properties and processes that control soil-moisture dynamics can be categorized as either dynamical or aggregation effects [16]. Dynamical effects of heterogeneity occur when land-surface heterogeneity induces processes (e.g. atmospheric circulations) that are not represented explicitly in a model. The aggregation effects of heterogeneity arise from spatially varying contributions to the water balance and surface atmosphere exchanges at scales smaller than a model's resolution. The discussions presented herein investigate the aggregation effects of heterogeneity at spatial scales up to the scale of a regional land-surface model or the footprint of a microwave remote sensor.

Land-surface models generally account for the aggregation effects by representing subgrid-scale heterogeneity either with statistical-dynamic or discrete methods [16, 56]. The statistical-dynamic methods define subgrid-scale spatial variability for a given variable $\theta$ over an area using an analytical or empirical probability density function, $f_{\mathrm{pdf}}(\theta)$, multiply all terms in the relevant grid-scale equations by $f_{\mathrm{pdf}}(\theta)d\theta$, and integrate over $f_{\mathrm{pdf}}(\theta)$ to obtain an area-averaged value of $\theta$ [16]. The simplest discrete method assumes that the land surface of a grid block is homogeneous with regard to its inputs and parameters. Inputs are area-averaged values, while parameters are a weighted average of all land-surface parameters or the most common land-surface parameters in the grid block. This discrete method is an 'effective-parameter' or 'lumped-parameter' approach, because a model developed and tested over a small homogeneous area is used at larger scales by redefining the parameters [24, 58]. Another discrete method is the mosaic approach, which divides a grid block into homogeneous subgrid blocks that directly and independently exchange fluxes with the atmosphere [4, 26]. The mosaic approach is considered an explicit method, because actual patterns of spatial variability are represented.

Land-surface models might use an effective-parameter approach to deal with one type of heterogeneity, while using a different approach for heterogeneity in another parameter or variable. Land-surface models that generally use only discrete methods include the biosphere-atmosphere transfer scheme (BATS) [12, 60], the land-surface model (Model II-LS) of the Goddard Institute for Space Studies [45], the simple biosphere models (SiB, SSiB, and SiB2) [48, 49, 59], the interaction soil biosphere atmosphere (ISBA) model [36], and the new set of parameterizations of the hydrologic exchanges (SECHIBA) [14]. The two-layer variable infiltration capacity (VIC-2L) model [30] and the parameterization for land-atmosphere-cloud exchange (PLACE) model [56] can represent heterogeneity using both statistical-dynamic and discrete methods.

We classify spatial heterogeneity as it relates to soil-moisture dynamics into two types: subgrid-scale variability of meteorological conditions and of the land surface [29]. For example, spatial heterogeneity of rainfall has been the focus of numerous studies. Liang et al. [29] developed a statistical-dynamic approach to include the spatial heterogeneity of rainfall. These authors evaluated the approach in terms of its ability to predict spatially averaged surface fluxes, runoff, and soil moisture produced by an explicit model. Sivapalan et al. [50] highlighted the profound effect of the subgrid-scale variability of rainfall on land-surface fluxes and on the long-term water balance. They demonstrated that spatially heterogeneous rainfall produces variability in the constitutive relationships for evapotranspiration and runoff and noted that the variability is particularly important for runoff predictions.

With regard to the subgrid-scale heterogeneity of the land surface, we distinguish between two types of spatial heterogeneity for modeling purposes [62]. The first type is vertical heterogeneity of the land surface, which includes variations of properties and variables in the vertical direction. Guswa et al. [19, 20] and Celia and Guswa [8] observed that vertical averaging of non-uniform properties and variables results in non-unique constitutive relationships between spatially averaged variables. Guswa [18] proposed a simple multi-valued relationship between transpiration and the vertically averaged soil-moisture content of the root zone to approximate the non-unique relationship. Non-uniqueness also arises in the relationships among soil-moisture contents averaged over different depths as demonstrated by Puma et al. [38]. These non-unique relationships have important implications for applications of soil-moisture remote sensing.

The second type of spatial land-surface heterogeneity has variations of relevant parameters and variables in the horizontal plane. Koster and Suarez [26] used the discrete mosaic approach to model different vegetation types that cover the land surface within a grid block. Bonan et al. [5] investigated the subgrid-scale variability of leaf area index, stomatal resistance, and soil moisture and found that predictions of latent heat flux and evapotranspiration were more sensitive than surface radiative flux to subgrid-scale heterogeneity. Zhu and Mohanty [62] examined vertical steady-state flow in a horizontally heterogeneous soil formation to derive and evaluate effective hydraulic parameters for an equivalent homogeneous medium. The analysis was extend by Zhu and Mohanty [61] to understand the effect of spatially variable root-water uptake on the effective hydraulic parameters. Nordbotten et al. [37] presented a probabilistic framework for characterization of spatially averaged evapotranspiration over an area with heterogeneous vegetation and uniform rainfall.

The main objective of this paper is to investigate the non-unique, spatially averaged constitutive relationships that arise due to spatial heterogeneity in rainfall and vegetation. In particular, we are interested in the scaling behavior of these constitutive relationships and the importance of the non-uniqueness for predicting components of the water balance. The analysis presented herein is accomplished using a discrete mosaic approach to represent spatial heterogeneity of rainfall and vegetation for a Texas water-limited ecosystem. In such ecosystems, rainfall and vegetation heterogeneity have a dominant control on the daily dynamics of soil moisture and evapotranspiration. Daily soil-moisture dynamics are less important in mesic and hydric ecosystems, since evapotranspiration is typically limited by atmospheric demand rather than soil moisture.

We first consider spatial heterogeneity in vegetation with spatially uniform rainfall by explicitly calculating the temporal evolution of soil moisture, evapotranspiration, and leakage. We examine the spatially averaged relationships of both evapotranspiration and leakage with soil moisture, and explore how spatial scales of heterogeneity affect the non-unique relationships. Next, spatially heterogeneity in rainfall is introduced, and we evaluate the soil-moisture-controlled relationships for various averaging areas. Soil-moisture and evapotranspiration predictions from the explicit mosaic approach are also compared with predictions from a simple effective-parameter approach, which takes a weighted average of land-surface parameters within the modeled grid block. Furthermore, we present a methodology to upscale soil-moisture-dependent functional relationships controlling the soil-water balance and discuss it in the context of the statistical-dynamic approach and the method proposed by Montaldo and Albertson [33].

## 2. Methodology

The simulations in this study represent the daily soil-moisture dynamics of a vegetated land surface for a single grid block in a regional-scale model. Properties of the land surface are relatively homogeneous at the scale of a single plant, such that the physical relationships controlling soil-moisture dynamics are taken to be well-defined. If plant-scale relationships are nonlinear and heterogeneity in land-surface properties and variables exists, then it is generally inappropriate to apply the plant-scale relationships at large spatial scales (e.g., [10, 37]).

We approach the spatially averaged relationships from the perspective of upscaling plant-scale relationships controlling soil-moisture dynamics to the grid-scale of a land-surface model. We intend upscaling to mean the derivation of relationships that involve averaged variables defined over relatively large spatial scales, based on known relationships for variables defined over smaller scales [8]. Although the focus of the upscaling is on the spatial scale corresponding to regional land-surface models, the results are relevant for general circulation models as well. In particular, we simulate and analyze the following scenarios:

1. *Spatially uniform rainfall with spatially heterogeneous vegetation;*
2. *Spatially varying rainfall with spatially homogenous vegetation; and*
3. *Spatially heterogeneous rainfall and vegetation.*

Subgrid-scale heterogeneity of the soil, plant, and atmosphere system is represented using a discrete mosaic approach, an effective-parameter approach, and an upscaled approach. We take the output from the mosaic approach to represent actual field conditions.

### *2.1. Spatial averaging*

Spatial averaging of variables must be performed in a manner that preserves their physical meaning. This analysis focuses on evapotranspiration, leakage, and soil moisture. Evapotranspiration (*ET*) is defined as volume of water per land-surface area per time that enters the atmosphere from the soil and vegetation. Leakage (*L*) is defined to represent the volume of water per land-surface area per time that leaves the root zone through the bottom boundary. As an example, spatially averaged evapotranspiration (<*ET*>) is expressed as an areally averaged variable in the discrete form

$$\langle ET \rangle = \frac{\sum_i A^i ET^i}{\sum_i A^i} \qquad (1)$$

where $A^i$ and $ET^i$ are the area and evapotranspiration of subgrid block *i*, respectively. Spatially averaged leakage is defined in an analogous manner. Conversely, relative soil-moisture content is defined as the volume of water per volume of voids in the soil, so that it is a volume-averaged variable given by

$$\langle S \rangle = \frac{\sum_i \phi^i Z_r^i A^i S^i}{\sum_i \phi^i Z_r^i A^i} \qquad (2)$$

where $\phi_i$ , $Z_r^i$, and $S^i$ are the porosity [$L^3$ voids/$L^3$ soil], depth of root zone [L], and relative soil-moisture content ($0 \le S^i \le 1$) [$L^3$ water/$L^3$ voids] of subgrid block *i*, respectively. That is, in order to ensure that the average quantities are physically meaningful, the areally defined variables are computed via area averaging, and volume defined variables are computed via volume averaging [22].

### *2.2. Discrete model using the mosaic approach*

The strategy for spatial discretization of the land surface depends on the processes modeled and the types of heterogeneity present. In fact, spatial discretization is the main factor that differentiates various models that use the mosaic approach [16]. For the case of spatially uniform rainfall and heterogeneous vegetation cover, we only need to simulate one subgrid block for each vegetation type, since the model consists of independent soil-water balances in horizontal space. The spatial average of variables will simply be a weighted area or volume average based on the fractional cover of each vegetation type. When spatial rainfall heterogeneity is modeled, the land surface is divided to simulate each plant (or homogeneous vegetation patch) since rainfall varies spatially in a random manner.

#### *2.2.1. Water balance at the plant-scale*

The model used to simulate soil-moisture dynamics in each subgrid block of the mosaic is based on a soil-water balance at a point and is expressed as [28]

$$\phi^i Z_r^i \frac{dS^i(t)}{dt} = R^i(t) - I^i\left[R^i(t)\right] - Q^i\left[R^i(t), S^i(t)\right] - ET^i\left[S^i(t)\right] - L^i\left[S^i(t)\right] \qquad (3)$$

where $R(t)$ is the rainfall rate [$L^3$ water/$L^2$ soil/T], $I[R(t)]$ is the amount of rainfall lost through interception [$L^3$ water/$L^2$ soil/T], $Q[R(t),S(t)]$ is the runoff rate [$L^3$ water/$L^2$ soil/T], $ET[S(t)]$ is the evapotranspiration rate [$L^3$ water/$L^2$ soil/T], $L[S(t)]$ is the leakage rate, and $t$ is time [T]. Superscript $i$ denotes the subgrid block. We assume that the root zones are non-overlapping in space and that they do not interact. For simplicity of notation, we will not indicate the time dependence of soil moisture in what follows unless it is necessary.

*2.2.2. Temporal and spatial rainfall model*

Rainfall input is treated as an external random forcing that is independent of soil moisture. The occurrence of storm events is modeled as a series of point events in time that arise according to a Poisson process with rate $\lambda_t$ (e.g., [28]). That is, we ignore the temporal structure within each storm event and apply each storm as a Dirac delta function in time. For simulations with spatially uniform rainfall, we obtain the rainfall depth from an exponential distribution with mean depth $E[Y]$.

When we explicitly consider spatial heterogeneity of rainfall, a storm event is modeled as spatially varying pulses of rainfall that represent daily precipitation. The spatial distribution of rainfall for a storm event is based on a simplified statistical description of the cellular structure of a storm event following Rodriguez-Iturbe et al. [39]. That is, a storm event is represented by a collection of rain-producing cells with each cell characterized by total depth of rainfall at the cell center and by a spread function, which specifies the decay of rainfall depth with distance from the cell center [39].

The cell centers are distributed over the region $R$ in a two-dimensional Poisson process of density $\lambda_{xy}$. We obtain the depth of rainfall at a cell center from an exponential distribution with mean rainfall depth $E[h]$. The total rain depth deposited in subgrid block $i$ ($Y^i$) is the sum of contributions from all cells in the region

$$Y^i = \sum_{(x^j, y^j) \in R} h^j \cdot g(r) \tag{4}$$

where $h^j$ is the rainfall depth at cell center $j$ due to cell $j$, $g(r)$ is a spread function, $r = \|(x^i, y^i) - (x^j, y^j)\|$ is the distance between $(x^i, y^i)$, the center of subgrid block $i$, and $(x^j, y^j)$, the center of the rain cell $j$. We model the spread of rainfall around cell centers according to a quadratic exponential function [39]

$$g(r) = \exp\left(-2\left(\frac{r}{a'}\right)^2\right) \tag{5}$$

where $a'$ [L] represents a characteristic spatial scale of a rain cell. The moments of this model are [39]

$$E[Y] = \frac{1}{2} \cdot \pi \lambda_{xy} (a')^2 E[h] \tag{6}$$

$$\mathrm{Var}[Y] = \frac{1}{2} \cdot \pi \lambda_{xy} (a')^2 E^2[h] \tag{7}$$

For the simulations in this paper, we assume that $a'$ is known, and we obtain $E[Y]$ and Var$[Y]$ from rainfall data. We then solve for $\lambda_{xy}$ and $E[h]$ by combining Equations (6) and (7).

### *2.2.3. Plant-scale interception and runoff*

Plant canopies intercept a significant portion of rainfall, especially in arid and semi-arid ecosystems where rainfall duration is short and evaporation demand is high [13, 28]. Following the simplified approach of Laio et al. [28], canopy interception is modeled by setting a fixed threshold rainfall depth, $\Delta$:

$$I^i\left[R^i(t)\right] = \min\left(\Delta^i, R^i(t)\right) \tag{8}$$

For rainfall depths greater than $\Delta^i$, interception is equal to $\Delta^i$. If a simulated storm produces a rainfall depth in subgrid block $i$ less than $\Delta^i$, then all of the rainfall is intercepted. For runoff, we have

$$Q^i\left[R^i(t), S^i(t)\right] = \max\left[0, R^i(t) - \phi^i Z_r^i\left(1 - S^i\left(t^-\right)\right)\right] \tag{9}$$

where $S^i(t^-)$ is the relative soil-moisture content at an instant before time $t$ in subgrid block $i$. That is, runoff only occurs if the water-storage capacity of the soil is exceeded; there is no Hortonian (infiltration excess) mechanism for runoff.

### *2.2.4. Plant-scale evapotranspiration*

For water-limited ecosystems, soil moisture has a dominant control on evaporation and transpiration. When the soil-moisture content of a plant's root zone is not sufficient to permit the normal course of plant physiological processes, transpiration is controlled by soil moisture (e.g., [28]). This plant-scale relationship between transpiration and soil moisture is well approximated by various deterministic functions as summarized and compared by Mahfouf et al. [32]. The differences among the functions are a result of diverse simplifying assumptions made to represent the complex soil-plant-atmosphere systems numerically [32]. Since we focus on vegetated land surfaces where evaporative fluxes are minor relative to transpiration fluxes (except when the soil is very dry), we model soil-water loss due to evaporation (excluding intercepted rainfall) and transpiration together as evapotranspiration.

A piecewise-linear function is used to approximate the plant-scale evapotranspiration relationship and is based on experimental evidence that demonstrates soil-moisture limitation of leaf conductance and stand transpiration [17, 28, 46]. A recent field investigation by Williams and Albertson [57] demonstrated that this simplified evapotranspiration function can be used successfully to model daily evapotranspiration and soil moisture at the plant scale in a savanna ecosystem. This plant-scale function is described mathematically as (e.g. [28, 44])

$$ET^i\left(S^i\right)=\begin{cases} E_w^i \dfrac{S^i - S_h^i}{S_w^i - S_h^i} & S_h^i < S^i \le S_w^i \\ E_w^i + \left(ET_{\max}^i - E_w^i\right)\dfrac{S^i - S_w^i}{S^{*,i} - S_w^i} & S_w^i < S^i \le S^{*,i} \\ ET_{\max}^i & S^* < S^i \le 1 \end{cases} \quad (10)$$

where $S_h$ is the relative soil-moisture content at the hygroscopic point, $S_w$ is the relative soil-moisture content at wilting, $S^*$ is the relative soil-moisture content at incipient stomatal closure, $E_w$ is the evaporation rate at $S_w$, $ET_{\max}$ is the maximum evapotranspiration rate, and superscript $i$ denotes the subgrid block. That is, if the soil moisture of the root zone is sufficiently high, then the evaporation rate is $ET_{\max}$, which is determined by atmospheric conditions and the characteristics of the plant. Plants close their stomata to prevent water loss when soil moisture is below the threshold soil-moisture value, $S^*$. The degree of stomatal closure within a plant's canopy increases as soil moisture decreases, so that the model reduces evapotranspiration linearly from $ET_{\max}$ to $E_w$ as the soil moisture decreases from $S^*$ to $S_w$. When the soil moisture reaches $S_w$, water is lost from the soil only through evaporation at very low rates until the soil reaches $S_h$ [28].

*2.2.5. Plant-scale leakage*

The rate of leakage from the root zone is assumed to be at a maximum when $S$ is equal to one and rapidly diminishes as the soil dries [28]. We assume that the maximum rate is equal to the saturated hydraulic conductivity, $K_s$ [$L^3$ water/$L^2$ soil/T], and that leakage is zero when the soil moisture reaches field capacity, $S_{fc}$ [28]. We further assume that the effective hydraulic conductivity decays exponentially, which lead to as the expression [28]

$$L^i\left(S^i\right)=\frac{K_s^i}{e^{\left(2b^i+4\right)\left(1-S_{fc}^i\right)}-1}\left[e^{\left(2b^i+4\right)\left(S^i-S_{fc}^i\right)}-1\right],\quad S_{fc}^i < S^i \le 1 \quad (11)$$

where $b^i$ is an experimentally determined parameter characterizing the soil in subgrid block $i$.

*2.2.6. Solution*

Because the rainfall input occurs instantaneously, Equation (3) is evaluated during inter-storm periods when the soil is drying. Water is lost from the soil only through evapotranspiration and leakage during these inter-storm periods, such that Equation (3) becomes

$$\frac{dS^i}{dt}=-\frac{\Phi^i\left(S^i\right)}{\phi^i Z_r^i}=-\frac{ET^i\left(S^i\right)+L^i\left(S^i\right)}{\phi^i Z_r^i} \quad (12)$$

where $\Phi^i(S^i)$ is referred to as the inter-storm soil-water loss function. Laio et al. [28] solved Equation (12) to obtain analytical expressions for soil-moisture decay. When rainfall occurs, canopy interception is calculated according to Equation (8), $S$ is increased, and runoff is computed according to Equation (9). Instantaneous values of *ET* and *L* are calculated using Equations (10) and (11), and the soil then dries over a day according to the analytical soil-drying expressions. The model must be interpreted at the daily timescale, because we ignore the temporal structure within each storm event and do not model the diurnal variation of $ET_{\max}$. Accordingly, the *S*, *ET*, and *L* predictions are instantaneous values with an interval of a day between predictions. It is important to note that the daily-integrated values of leakage will be significantly less than the instantaneous values reported. Spatially averaged variables are obtained following the averaging procedures described in Section 2.1.

### *2.3. Discrete methods using implicit approaches*

We also simulate the regional-scale grid block using two methods, which implicitly account for spatial heterogeneity of the land surface. The first method is a common effective-parameter approach, where the effective parameters are calculated via weighted averages of the subgrid parameter values. The second approach obtains upscaled functions based on predictions from the explicit mosaic approach. In this latter method, we do not attempt to determine the relationships a priori, but rather explore the effects of using a one-to-one relationship when the true behavior is non-unique.

#### *2.3.1. Effective-parameter approach*

In the effective-parameter approach, the soil-water balance equation (Eq. (3)) and its constitutive equations (Eq. (8) – (11)) are applied at a single point for the entire grid block. That is, an 'average' vegetation type covers the entire grid box. Spatially heterogeneous rainfall generated by the spatial rainfall model is areally averaged and used as hydrologic input. Parameters are weighted averages of the parameters of plants present in the grid block. As with spatially averaged variables, parameters defined per unit area are spatially averaged using area averaging, while parameters defined per unit volume are spatially averaged using volume averaging. We use this approach to simulate the three scenarios considered with the mosaic approach and compare predictions of <*ET*>, <*L*>, and <*S*>.

#### *2.3.2. Upscaled-parameter approach*

The upscaled-parameter approach accounts for heterogeneity by replacing plant-scale functional relationships with relationships that can be applied at the scale of a model's grid block. As with the effective-parameter approach, areally averaged rainfall is the hydrologic input. The goal is to upscale the inter-storm soil-water loss function to improve predictions of <*ET*> and <*S*>. The daily output of <*ET*>, <*S*>, and <*L*> from the explicit mosaic approach are analyzed to approximate the relationships between <*ET + L*> and <*S*> and between <*ET*> and <*S*> for the case with spatially heterogeneous rainfall and vegetation. In particular, we solve the equation

$$\frac{d\langle S\rangle}{dt} = -\frac{\Phi_{\mathrm{grid}}\left(\langle S\rangle\right)}{\langle \phi\rangle\langle Z_r\rangle} \qquad (13)$$

where $\Phi_{\text{grid}}(\langle S\rangle)$ is a grid-scale inter-storm soil-water loss function. $\Phi_{\text{grid}}(\langle S\rangle)$ is approximated using the method of least squares to fit the relationship between ($\langle ET\rangle + \langle L\rangle$) and $\langle S\rangle$ and is given by

$$\Phi_{\text{grid}}\left(\langle S\rangle\right)=\begin{cases} 0 & 0<\langle S\rangle\le\langle S_h\rangle \\ \langle E_w\rangle\cdot\dfrac{\langle S\rangle-\langle S_h\rangle}{\langle S_w\rangle-\langle S_h\rangle} & \langle S_h\rangle<\langle S\rangle\le\langle S_w\rangle \\ c_1\langle S\rangle+c_2 & \langle S_w\rangle<\langle S\rangle\le\langle S_{fc}\rangle \\ c_3\langle S\rangle+c_4 & \langle S_{fc}\rangle<\langle S\rangle\le 1 \end{cases} \tag{14}$$

where $c_1$, $c_2$, $c_3$, and $c_4$ are empirical parameters. We are also interested in improving $\langle ET\rangle$ predictions, so we approximate a grid-scale evapotranspiration function using a functional form motivated by our empirical results:

$$ET_{\text{grid}}\left(\langle S\rangle\right)=\begin{cases} \langle E_w\rangle\cdot\dfrac{\langle S\rangle-\langle S_h\rangle}{\langle S_w\rangle-\langle S_h\rangle} & \langle S_h\rangle<\langle S\rangle\le\langle S_w\rangle \\ \langle ET_{\max}\rangle\cdot\left[1-\left(c_5-c_6\langle S\rangle\right)^2\right] & \langle S_w\rangle<\langle S\rangle\le\langle S_{fc}\rangle \\ \langle ET_{\max}\rangle & \langle S_{fc}\rangle<\langle S\rangle\le 1 \end{cases} \tag{15}$$

where $c_5$ and $c_6$ are parameters. The analytical expression for soil-moisture decay based on Equation (13) from an initial condition $\langle S_0\rangle$ in the absence of rainfall is given in the Appendix. Note that for $\langle S\rangle$ values between $\langle S_h\rangle$ and $\langle S_w\rangle$, we apply the weighted average approach. Since the soil seldom dries to these low values, slight modifications to the function for this range of $\langle S\rangle$ does not have a significant effect on predictions of $\langle ET\rangle$ and $\langle S\rangle$. We evaluate this upscaling approach through comparison with predictions from the mosaic and the effective-parameter approaches.

## 3. Simulation of a water-limited Texas ecosystem

The ecosystem examined in this analysis is a savanna parkland/woodland vegetation complex located in the eastern Rio Grande Plains of Texas. The region has been extensively studied by the Texas Agriculture Experiment Station in the *La Copita Research Area*, Texas (27°40'N, 98°12'W). The land surface is characterized by a flat landscape with very mild slopes that is covered by a diphase tree-grass vegetation [27, 47]. Although the ecosystem's potential natural vegetation is classified as *Prosopis-Acacia-Andropogon-Setaria* savanna, a shift from grass to woody plant domination has occurred within the past 150 years (e.g., [2]). *Prosopis glandulosa* (honey mesquite) is the dominant woody plant (e.g. [2, 3, 7, 47]), which coexists with $C_4$ grasses including *Paspalum setaceum* [3]. The properties of individual *Prosopis glandulosa* and *Paspalum setaceum* are presented in Table 1 along with parameters that describe the vegetation used in the effective-parameter approach.

The eastern Rio Grande Plains, a water-limited ecosystem, has approximately 70% of its annual rainfall occurring from April to September [27, 47]. The data for this site consist of daily precipitation records for 20 years (1977 to 1997, excluding 1982), obtained from the National Climatic Data Center (NCDC) station at Benavides, Texas, which is located near the Texas Agriculture Experimental Station. For this analysis, we consider only the period from May 15 to June 16 to obtain a rainfall record that is approximately homogeneous (statistically) in time, because temporal rainfall statistics change during the growing season [40-42]. Simulations are run for 20,000 days, so that we can analyze the statistics of variables controlling soil-moisture dynamics. We assign a value to the characteristic size of a rain cell (as captured by parameter $a'$ ) that is consistent with observation of thunderstorms [35, 51]. $E[Y]$ and $\mathrm{Var}[Y]$ are estimated using the NCDC daily precipitation record, while $\lambda_{xy}$ and $E[h]$ are then calculated using Equations (6) and (7). These rainfall model parameters are given in Table 2.

The soil in La Copita is spatially heterogeneous. Approximately 57% of the site has an A horizon of fine sandy loam, which is uniformly distributed over the area [47]. To keep the focus on spatial heterogeneity in rainfall and vegetation, we make the simplification that the soil of the root zone is homogeneous. Table 3 presents the physical parameters of the soil used in the simulations from Laio et al. [27]. When we investigate the effects of spatial vegetation heterogeneity, the root-zone depth, $Z_r$, varies spatially.

We simulate two land-surface types. The first type is a grassland, so that we can analyze the effects of spatial rainfall heterogeneity separately. The second land-surface type represents the spatial structure of vegetation as a matrix of grass with woody plants located according to a Poisson process in space (e.g., [43]). The woody plants are assumed to have circular crowns, where the radius of each crown is obtained from an exponential distribution. We discretize the land surface into 5 m×5 m subgrid blocks, which have an area approximately equal to the average canopy area of the woody species *Prosopis glandulosa*. Mature *Prosopis glandulosa* in southern Texas typically have canopy diameters of 6.5 m or less, depending on the characteristic of the soil [2]. For simulation purposes, the soil-moisture balance is calculated for a sparse sample of subgrid blocks, which has been verified to reproduce predictions of spatially averaged variables from the full sample.

The inter-storm soil-water loss functions for the grass *Paspalum setaceum* and the woody plant *Prosopis glandulosa*, characterized by the parameters in Table 1, are presented in Fig. 1a. The functions are very similar in terms of parameter values for $S_w$, $S^*$, $E_{max}$, and $ET_{max}$, but $Z_r$ is significantly different for the two vegetation types. Importantly though, the value of $\phi Z_r$ for the grass is less than half of the $\phi Z_r$ for the woody plant. Consequently, differences in soil-moisture and evapotranspiration predictions between the two vegetation types are primarily due to the parameter $Z_r$.

## 4. Results and discussion

The results that follow demonstrate the complexity of upscaling inter-storm soil-water loss and evapotranspiration as functions of soil moisture from the plant scale to the regional scale. Celia and Guswa [8] considered whether, in general, hysteresis should be expected *a priori* whenever upscaling is performed for non-linear equations. Hysteresis, in the context of this paper, simply refers to history dependence of upscaled or spatially averaged relationships. A consequence of hysteresis is the existence of non-uniqueness between variables. The following results demonstrate that non-unique relationships exist for all three of the spatial heterogeneity scenarios simulated.

### 4.1. *Spatially uniform rainfall with spatial heterogeneity in vegetation type*

#### *4.1.1. Non-uniqueness in spatially averaged relationships*

Fig. 1b compares spatially averaged inter-storm soil-water loss, <*ET*> + <*L*>, as a function of <*S*> from the mosaic approach with the effective-parameter function, when rainfall input is equal for both simulations. Each marker on the plot corresponds to a single set of daily values of the variables. A non-unique relationship exists between the spatially averaged variables from the mosaic approach, even though unique relationships are applied locally. At high values of <*S*> (e.g. above ~0.5), the mosaic approach predicts the same <*ET*> as the effective-parameter function. This result for <*ET*> implies that <*L*> is often underestimated by the effective-parameter approach. That is, the effective-parameter approach predicts that leakage occurs only if <*S*> is greater than <$S_{fc}$> (0.56, in this case). However, the results from the explicit method demonstrate that leakage often occurs for lower values of <*S*> beginning at ~0.45.

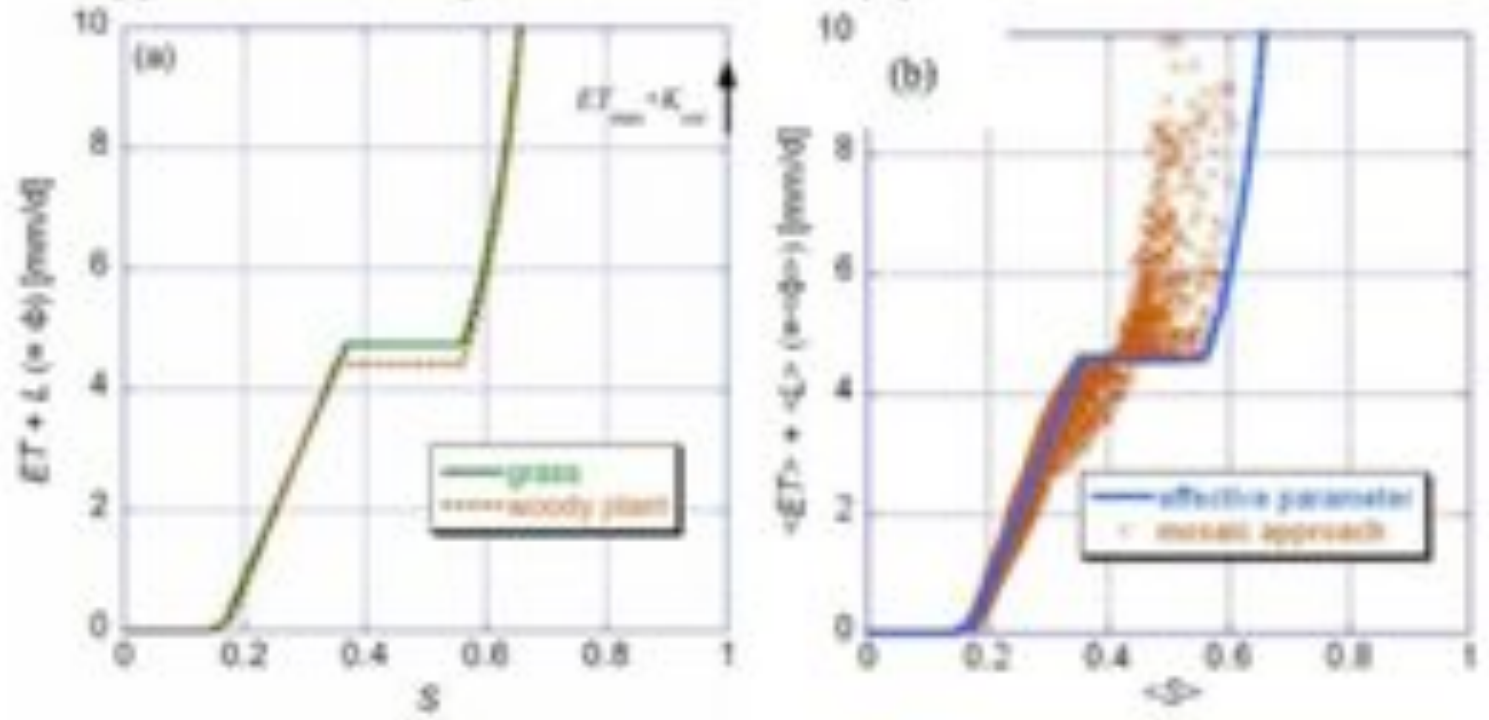

**Figure** 1: (a) Soil-water loss functions, Φ, at the plant scale for the grass *Paspalum setaceum* and the woody plant *Prosopis glandulosa* in a Texas water-limited ecosystem. (b) Comparison of <Φ> versus <*S*> as predicted by the effective-parameter approach and the mosaic approach for 50% grass and 50% woody plant with uniform rainfall.

#### *4.1.2. Region of accessible soil-moisture contents*

Fig. 2 is a scatter plot of the individual values of *S* for the grass and woody plant, which are subsequently averaged to obtain <*S*>, as simulated with the mosaic approach. Nordbotten et al. [37] discussed in detail the movement in this relative soil-moisture space in response to various sequences of infiltration and drying events. Based on the observations of Nordbotten et al. [37], we can define the bounds for the region of accessible values of soil moisture. The lower bound corresponds to the case where the initial soil moisture is at [$S_h$,$S_h$], and a rainfall event falls evenly over the grass and woody plant. The depth of rainfall that increases the soil-moisture content of the grass from $S_h$ to one is

$$h^{lb} = \phi Z_{r,g}\left(1 - S_h\right) + \Delta_g \quad (16)$$

where subscripts *g* denotes grass and superscript *lb* stands for lower bound. Since rainfall is spatially uniform, it increases soil moisture from any initial value in Fig. 2 with a constant slope equal to

$$m^{lb} = \frac{\phi Z_{r,g}\left(1 - S_h\right) + \Delta_g - \Delta_{wp}}{\phi Z_{r,wp}\left(1 - S_h\right)} \quad (17)$$

where subscript *wp* denotes woody plant. Therefore, the equation for the lower bound is given by

$$S_{wp}^{lb} = m^{lb}\left(S_g^{lb} - S_h\right) + S_h \quad (18)$$

The region has an upper bound given by a drying curve that originates at [1,1] and terminates at [$S_h$,$S_h$]. This drying curve is the upper bound, because its slope is always greater the slope of wetting events given in Equation (17) for the parameters used in these simulations. The non-uniqueness occurs because a given value of <*S*> can be obtained from combinations of grass and woody plant *S* within these bounds. An example of this region's usefulness is that we can compute a field-capacity value of <*S*> as the average of $S_{fc}$ for the grass and the corresponding tree *S* on the lower bound given by Equation (18).

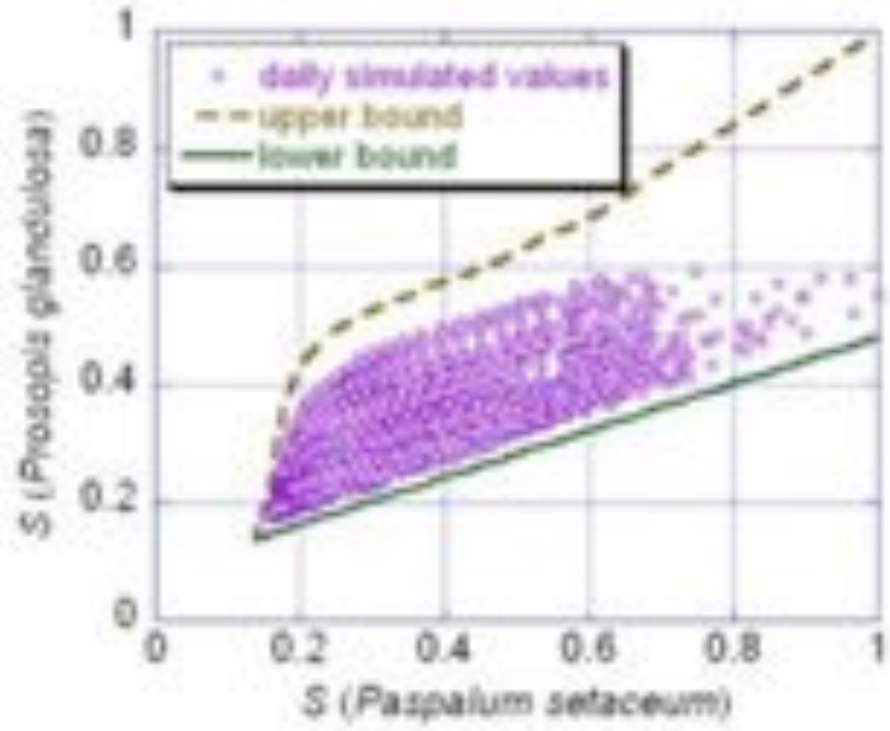

**Figure 2**: Scatter plot of *S* for the woody plant *Prosopis glandulosa* versus *S* for the grass *Paspalum setaceum* when spatially uniform rainfall is applied to the two plants. The region of possible relative soil-moisture contents is bounded by the specified curves.

### *4.1.3. Hysteresis between spatially averaged variables*

Fig. 3 shows a single wetting and drying event, and its effects on the relationships of <*ET*> and <Φ> with <*S*>. In Fig. 3a, we see that a wetting and drying event produces a counter-clockwise path in the <*ET*> versus <*S*> plot. Conversely, Figure 3b demonstrate that this same event generates clockwise paths in the <Φ> versus <*S*> space. The wetting and drying pathway will depend on the combination of *S* values producing the initial <*S*> value. That is, the spatially averaged relationships exhibit hysteresis, and the hysteresis produces non-uniqueness between spatially averaged variables.

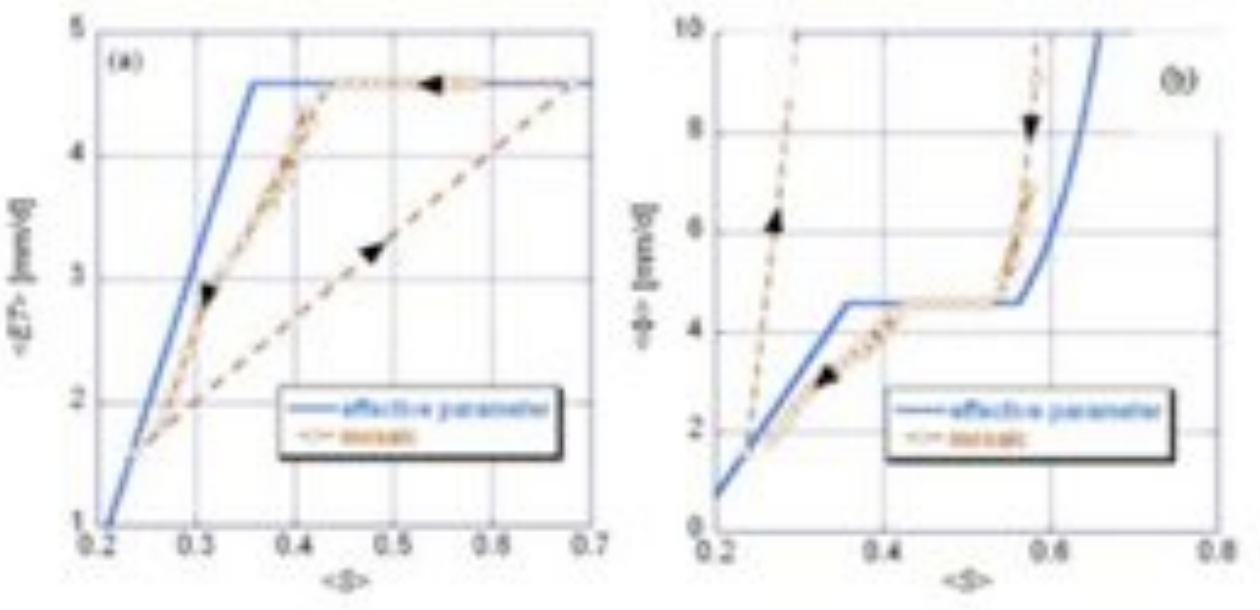

**Figure 3**: Temporal trace of wetting and drying sequence in the (a) <*ET*> versus <*S*> space, and (b) <Φ> versus <*S*> space for simulations of spatially uniform rainfall in a Texas ecosystem.

## *4.2. Spatial heterogeneity in rainfall with homogeneous vegetation cover*

The rainfall model produces thunderstorm rainfall, which is characterized by spatial heterogeneity of rainfall produced through the number and spatial location of rain cells and the distribution of rain within these cells. The heterogeneity within a grid block is also a function of time because of the random number, location, and depth of rain cells. We use the mean size of a rain cell within a storm, $E[A_{rain_cell}]$, to generalize our results with respect to spatial rainfall heterogeneity within thunderstorms. For our rainfall simulations, $E[A_{rain_cell}]$ is equal to 130 km$^2$. We present results for three different grid-block sizes ($A_{grid}$), 1 km$^2$ (field scale), 25 km$^2$, and 900 km$^2$ (regional scale), which correspond to $A_{grid}$ /$E[A_{rain_cell}]$ values of 0.008, 0.2, and 7.0. Since the vegetation cover is spatially uniform, the soil-water loss function of the effective-parameter approach is equivalent to the plant-level function for this analysis.

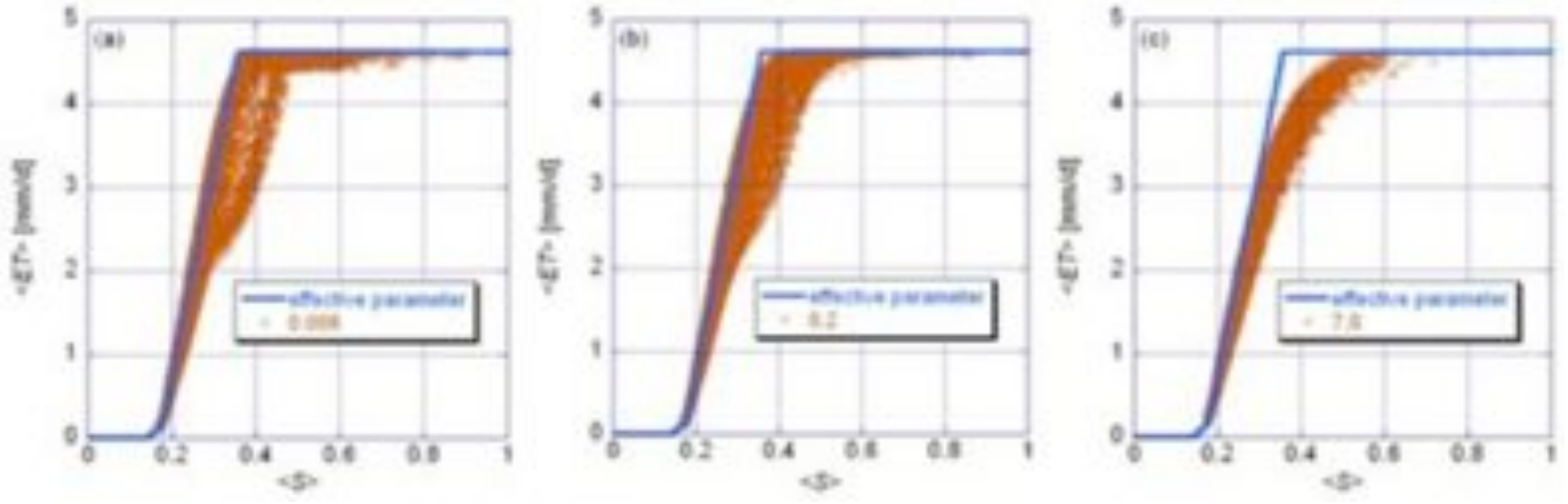

**Figure 4**: Relationships between <*ET*> versus <*S*> as predicted by the mosaic approach (markers) and the effective-parameter approach (solid line) when rainfall is spatially non-uniform for $A_{grid}/E[A_{storm}]$ equal to (a) 0.008, (b) 0.2, and (c) 7.0. The $r^2$ values are 0.96, 0.94, and 0.72, respectively.

A non-unique relationship is evident (Fig. 4) for $A_{grid}$ /$E[A_{rain_cell}]$ equal to 0.2, because rainfall is spatially heterogeneous due to random rain-cell locations and the rainfall distribution within a cell (Equation (5)). The effective-parameter approach overestimates <*ET*> for a large range of <*S*>, but it predicts <*ET*> values close to the means of the <*ET*> distributions for <*S*> values near <$S_w$> or above <$S_{fc}$>. We quantify the results of Fig. 4 using the coefficient of determination, $r^2$, as the goodness of fit of the effective-parameter approach. Even if the land surface is homogeneous, these results demonstrate that spatial heterogeneity in rainfall produces non-unique constitutive relationships between spatially averaged variables.

### *4.3. Spatial heterogeneity in rainfall and vegetation type*

When vegetation cover is non-uniform, two spatial scales are important as they relate to the averaging area for <*ET*> and <*S*> predictions: the scales associated with rainfall and vegetation heterogeneity. The rainfall heterogeneity is the same as for the previous simulations. We simulate a savanna with 56.6% grass cover and 43.4% woody plant cover by assigning the Poisson arrival rate of woody plants to be 0.04 $m^{-2}$ and the average crown radius to be 1.5 m. We do not have clustering of grass or woody plants with this vegetation structure, and the spatial distribution of vegetation does not change with time. Given this vegetation structure, spatial heterogeneity in vegetation generally occurs at spatial scales above 25 $m^2$ (the spatial resolution of the model).

#### *4.3.1. Spatial scale and non-uniqueness in spatially averaged relationships*

The distributions of <*ET*> for a given <*S*> are presented in Fig. 5 for the three values of $A_{grid}$ /E[$A_{rain_cell}$]. We find that the pdfs of <*ET*> for <*S*> equal to 0.3±0.0025 shift to the left and means of the distributions decrease as $A_{grid}$ /E[$A_{rain_cell}$] increases. The shift of the distributions is related to the range of plant-scale *S* values that produces this <*S*> value. This range becomes larger as $A_{grid}$ /E[$A_{rain_cell}$] increases, such that <*S*> is often the result of spatial averages including large *S* values (that is, above *S**). Because the plant-scale *ET* function is constant above *S**, <*ET*> values will be lower as more values of *S* above *S** are included in the spatial average.

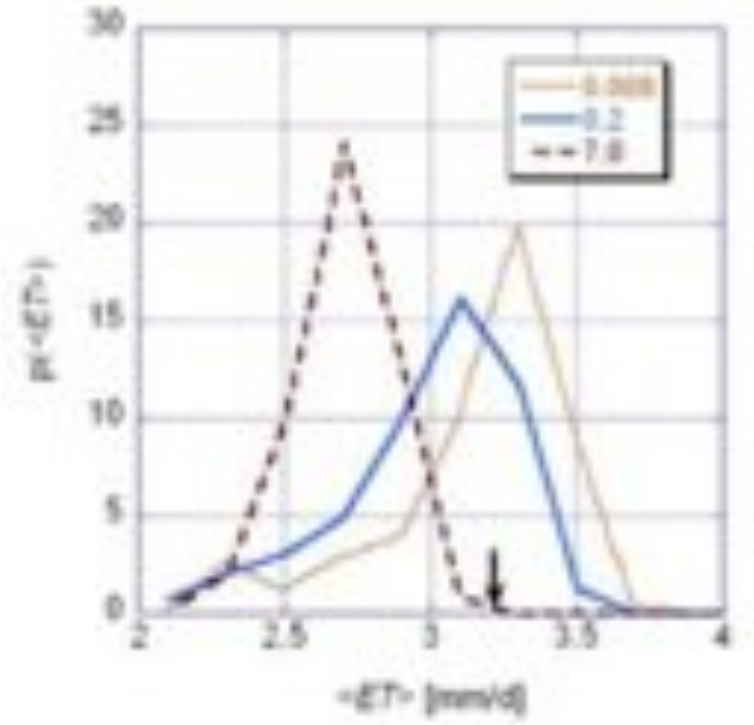

**Figure 5**: Comparison of <*ET*> pdfs for $A_{grid}$/E[$A_{storm}$] equal to 0.008, 0.2, and 7.0 when <*S*> is 0.3±0.0025. The arrows indicate the <*ET*> value predicted by the effective-parameter approach (3.2 mm/d).

Fig. 6 presents the relationship between daily <*L*> and <*S*> from the mosaic and effective-parameter approaches. That is, <*S*> is the instantaneous value at time *t*, and the daily <*L*> is the total leakage that occurs in the 24 hours after time *t*. The non-unique relationship between daily <*L*> and <*S*> is shown in Fig. 6a for $A_{grid}$ /E[$A_{rain_cell}$] equal to 0.008. The mosaic approach predicts that leakage often occurs for <*S*> values below <$S_{fc}$>. Fig. 6b shows the same relationship for $A_{grid}$ /E[$A_{rain_cell}$] equal to 7.0. Leakage occurs more frequently and at lower <$S$> values for the larger averaging area, with daily <$L$> predictions for given values of <$S$> also spanning several orders of magnitude. A considerable number of daily <$L$> values less than 1 mm/d occur, because leakage is highly localized. That is, daily $L$ values may be high for only a small area of grass within the averaging area, but daily <$L$> is a low because we average over the entire grid block. Some large-scale models, such as the ARNO model [53], attempt to account for leakage at low values of <$S$>. However, the ARNO model uses a deterministic empirical function, which is based on experience from application [53].

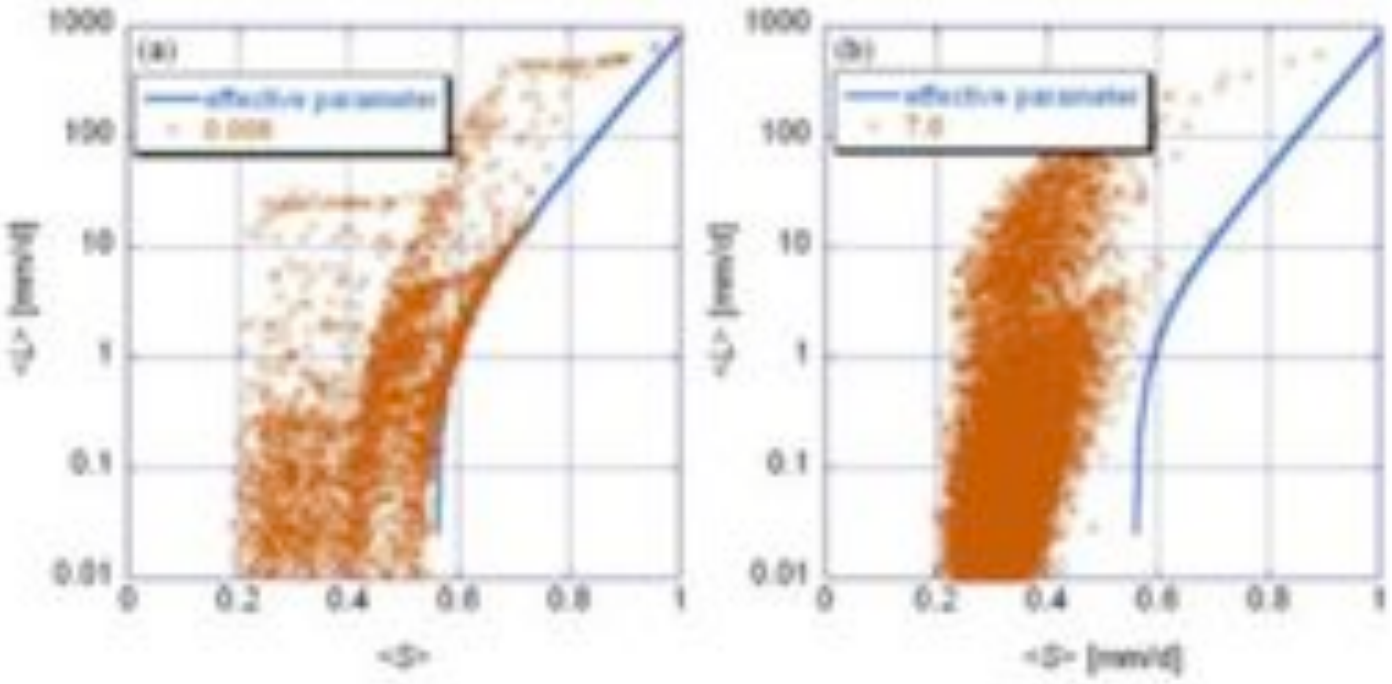

**Figure 6**: Relationships between <$L$> versus <$S$> as predicted by the mosaic approach (markers) and the effective-parameter approach (solid line) for $A_{grid}$/E[$A_{storm}$] equal to (a) 0.008 and (b) 7.0.

### *4.3.2. Upscaling functional relationships*

The upscaled functions given by Equations (14) and (15) are based on the mosaic-approach predictions of <$ET$> and <$L$>. We fit $\Phi_{grid}$ and $ET_{grid}$ for the case of spatially heterogeneous rainfall and vegetation when $A_{grid}$ /E[$A_{rain_cell}$] equals 7.0. The piecewise linear function given by Equation (14) is fit to the data as shown in Fig. 7a. The linear least-squares method was applied for <$S$> values between <$S_{fc}$> and 1. When fitting $\Phi_{grid}$ predictions corresponding to <$S$> values between <$S_w$> and <$S_{fc}$>, we find that the errors are not normally distributed. Consequently, a robust least-squares method with bisquare weights was used for the range <$S_w$> to <$S_{fc}$> [15]. Fig. 7b shows the upscaled $ET_{grid}$ function, which was fitted using the Levenberg-Marquardt method of non-linear least-squares regression [34]. The parameters calculated using these regression schemes are given in Table 4.

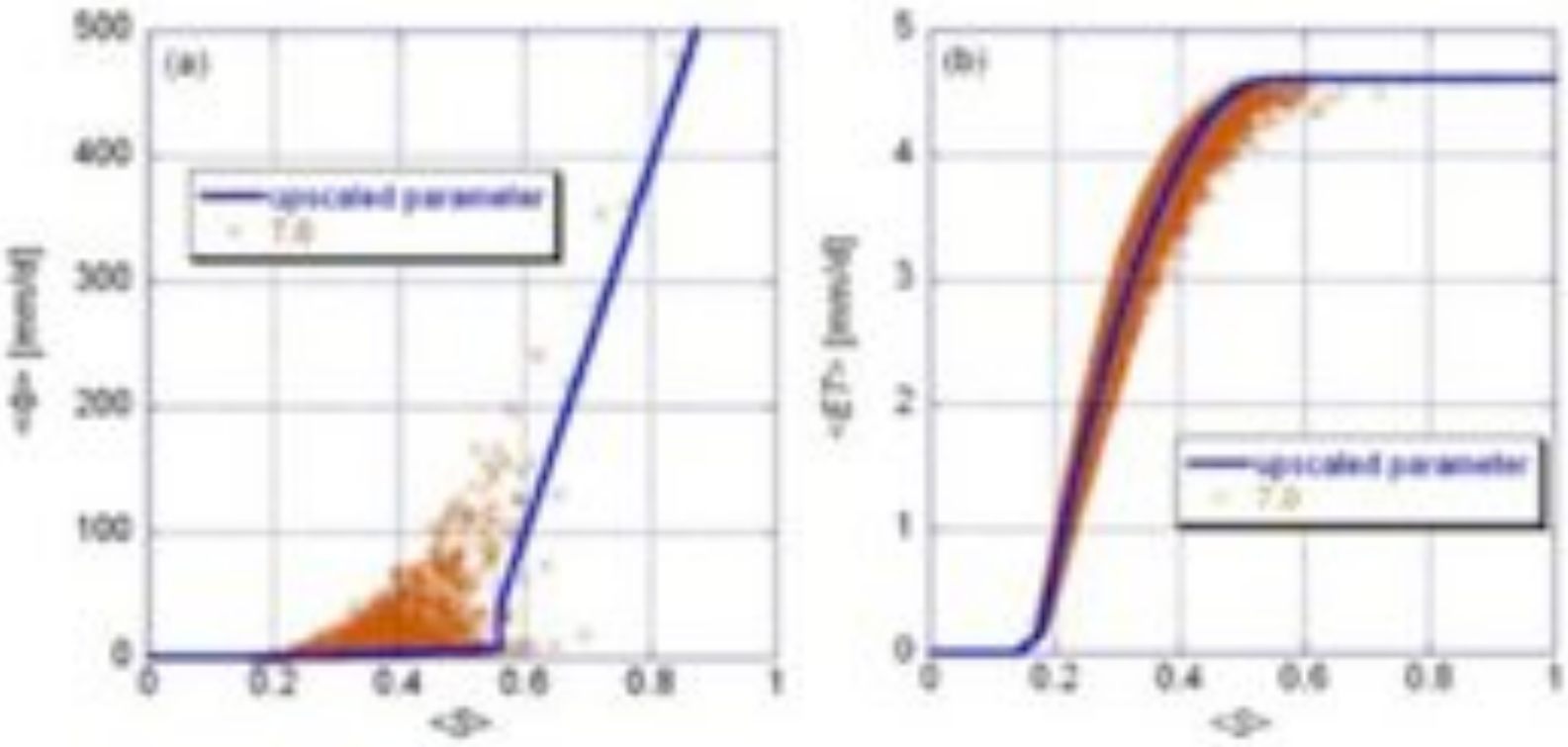

**Figure 7**: Upscaled functions for (a) the <$\Phi$> versus <$S$> relationship and (b) the <$ET$> versus <$S$> relationship for $A_{grid}$/E[$A_{storm}$] equal to 7.0.

### *4.3.3. Evaluation of the upscaled <S> and <ET> predictions*

Fig. 8a shows a portion of the 20,000-day time trace of <*S*> comparing predictions from the mosaic approach with the effective-parameter and upscaled approaches for $A_{\text{grid}}$ /E[$\text{A}_{\text{rain_cell}}$] equal to 7.0. Following large storm events, the upscaled approach improves <*S*> and <*ET*> predictions when compared with the effective-parameter approach. The differences are particularly pronounced in terms of <*ET*> predictions, as shown in Fig. 8b. We find that the mosaic, effective-parameter, and upscaled parameter approaches predict <*ET*>, averaged over the period from day 900 to day 1000, to be 3.39±0.83 mm/d, 4.13±0.72 mm/d, and 3.43±0.86 mm/d, respectively. During periods without large storm events, the upscaled parameter approach improves predictions but not as dramatically. As an example, we consider the first 300 days of the simulations, during which no large storm occurred (<*S*> did not exceed 0.41). The mosaic, effective-parameter, and upscaled parameter approaches predict <*ET*> averaged over the period to be 1.98±0.86 mm/d, 2.17±1.13 mm/d, and 2.11±0.92 mm/d, respectively.

Excluding periods of immediately after large rainfall events, both approaches are relatively accurate in terms of the timing and magnitude of <*S*> and <*ET*> predictions. This accuracy is first due to the areally averaged rainfall input. This rainfall input allows the effective-parameter approach to predict peaks in <*S*> that coincide with those predicted by the mosaic approach. A second reason for the accuracy of these two approaches is that evapotranspiration is a self-limiting process due to its negative feedback on soil moisture. For example, the effective-parameter approach might overestimate <*ET*> after a storm compared to the mosaic prediction. At a subsequent time, its predictions would improve, because the <*S*> value used to calculate <*ET*> would be lower than the mosaic prediction. This self-limiting nature of evapotranspiration also minimizes the importance of the non-uniqueness between <*ET*> and <*S*> as well as the differences between the effective-parameter and upscaled parameter approaches.

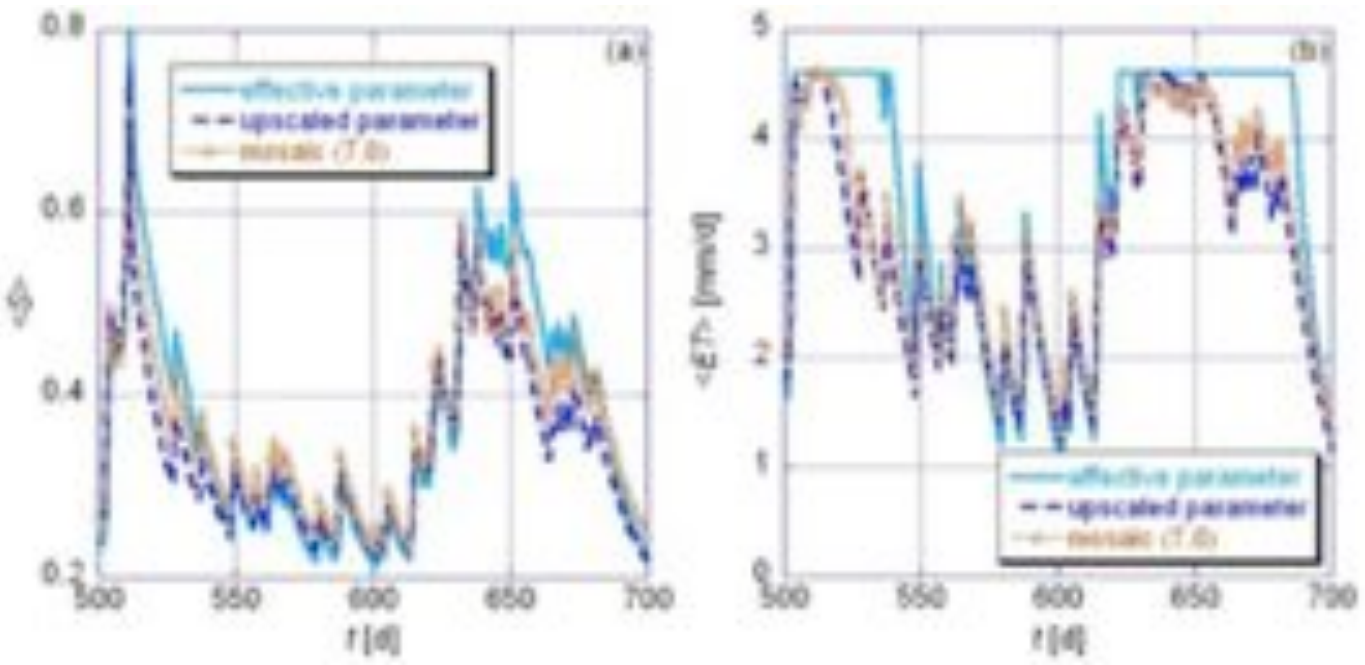

**Figure 8**: Temporal traces of (a) <*S*> and (b) <*ET*> comparing predictions from the mosaic approach with predictions from the effective-parameter approach and the upscaled approach for $A_{\text{grid}}$/E[$A_{\text{storm}}$] equal to 7.0.

If we consider the water balance for the 20,000-day simulation, we find that the mosaic approach predicts that 90.2% of infiltrating rainfall goes as evapotranspiration, 8.4% as leakage and 1.4% as runoff. The partitioning of infiltrating rainfall is 99% evapotranspiration, 1% leakage, and no runoff for the effective-parameter approach and 93% evapotranspiration, 7% leakage, and no runoff for the upscaled parameter approach. Clearly, the upscaled parameter approach improves the long-term temporal average of $<L>$ when compared to the effective-parameter approach. However, if we consider the mean daily differences between the mosaic approach and the effective-parameter and upscale parameter approaches, we find values equal to 2.3 mm/d and 2.1 mm/d, respectively. Clearly, the timing and magnitude of $<L>$ predictions is not significantly improved using the upscaled parameter approach. Consequently, if the goal is to predict $<L>$ at the daily timescale, single-valued functional relationships are not sufficient to approximate the non-unique relationships between $<L>$ and $<S>$.

Despite the imprecision of $<L>$ predictions, a single-valued function can predict $<S>$ and $<ET>$ accurately. Leakage, like evapotranspiration, is a self-limiting process, and, according to the plant-level leakage function, the rate of leakage decreases exponentially as the soil dries. After a storm event, leakage can exhibit significant spatial variability, but this variability will decrease rapidly given the short timescale of leakage dynamics. Furthermore, leakage acts to reduce the spatial variability of $S$, which is produced by spatial rainfall heterogeneity. Even though we fail to predict high rates of leakage locally, the soil quickly dries to $S_{fc}$ in these areas. In addition, the local inaccuracies in soil-moisture values immediately after a storm do not affect $<ET>$ predictions significantly, because $ET$ is independent of soil moisture above $S^*$. Therefore, a single-valued function combined with areally averaged rainfall input is sufficient to predict $<S>$ and $<ET>$ accurately.

Spatially averaged leakage is an important quantity to predict for many applications. As an example, $<L>$ is essential for estimating recharge of groundwater aquifers, especially in water-limited ecosystems such as La Copita, Texas. Our results demonstrate that a single-valued leakage function cannot be used with areally averaged rainfall input to predict $<L>$ accurately in the simulated ecosystem. The reason for the failure to predict $<L>$ accurately, in light of the success of predicted $<ET>$, is linked to the shape of the $L$ function and the short timescale of leakage dynamics. That is, the exponential form of the $L$ function predicts large changes in $L$ for small changes in $S$, relative to those for the $ET$ function. We also point out that the $ET$ function is constant (equal to $ET_{\max}$) for soil-moisture values above $S^*$. Finally, the maximum leakage rate, $K_{sat}$, is two orders of magnitude larger than $ET_{\max}$.

## 5. Other approaches and the relevance of the upscaled approach

### *5.1. Statistical dynamic approach*

The basic requirement of the statistical dynamic approach is knowledge of probability density functions (pdfs) for relevant variables to represent subgrid-scale spatial variability over an area. The general strategy in land-surface models has been to consider heterogeneity for only a few parameters and variables and to assume that they are independent [6, 16, 33]. In addition, the temporal evolution of pdfs of variables is typically ignored. One exception is the framework developed by Nordbotten et al. [37], which allows for the computation of the time varying pdfs of $<S>$ and $<ET>$ over an area with spatially heterogeneous vegetation and spatially uniform rainfall. The authors found that it is necessary to use a joint probability density function to characterize soil moisture for different vegetation types, because use of the marginal soil-moisture pdfs results in underestimation of <*ET*> for a given <*S*> over an extended range of soil-moisture values. The statistical dynamic approach has not been derived to include spatial heterogeneity in both rainfall and vegetation.

We focus on the case of spatial heterogeneity in rainfall with homogeneous vegetation cover (Section 4.2) to discuss the statistical dynamic approach. This case allows us to avoid the complexity of dependence among pdfs of *S* for multiple vegetation types. In particular, we are interested the relationship between the spatially averaged inter-storm soil-moisture loss, <Φ>, and <*S*> as predicted by the statistical dynamic approach. For the purposes of this discussion, we assume soil moisture is normally distributed due to the spatially heterogeneous rainfall (e.g. [10]). We can express the grid-scale inter-storm soil-water loss function, $\Phi_{\text{grid}}(\langle S \rangle)$, as

$$\Phi_{\text{grid}}\left(\langle S\rangle\right)=\int_0^1 \Phi(S)\, f\left(S;\langle S\rangle,\sigma_S^2\right)dS \quad (19)$$

where $f(S;\langle S\rangle,\sigma_S^2)$ is the normal distribution with <*S*> and the spatial variance of the daily *S* values, $\sigma_S^2$ , as parameters.

The spatial variance of *S*, $\sigma_S^2$, is dynamic in time as demonstrated by Albertson and Montaldo [1]. We use the mosaic approach to calculate the spatial variance of *S* for each day of our grassland simulations and plot $\sigma_S^2$ versus <*S*> for three values of $A_{\text{grid}}$ /E[$A_{\text{rain_cell}}$] in Fig. 9. These plots demonstrate that a highly non-unique relationship exists between $\sigma_S^2$ and <*S*> and that the characteristics of the relationship change depending on the scale of rainfall heterogeneity. It is necessary to approximate $\sigma_S^2$ for the statistical dynamic approach. The typical simplification is to assume that $\sigma_S^2$ is constant, thereby ignoring its temporal dynamics (e.g. [10]).

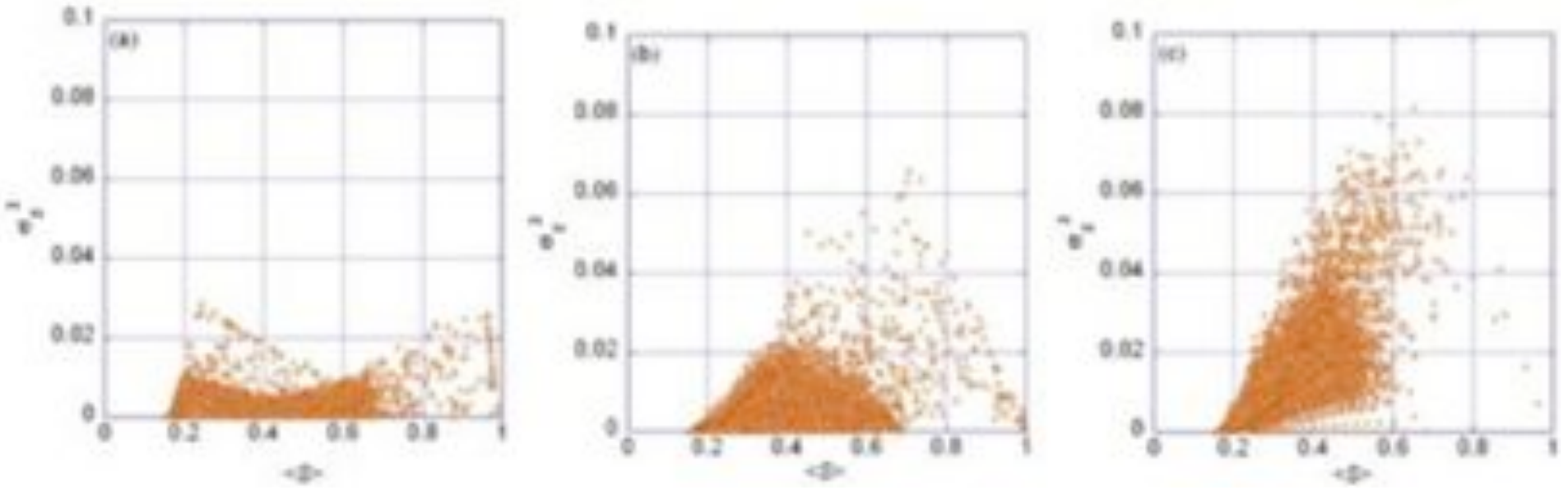

**Figure 9**: Plot of the spatial variance of daily *S*, $\sigma_S^2$, versus <*S*> for $A_{\text{grid}}$/E[$A_{\text{storm}}$] equal to (a) 0.008, (b) 0.2, and (c) 7.0 when rainfall is spatially non-uniform for a hypothetical Texas grassland.

Fig. 10 shows the relationship between <Φ> and <*S*> assuming that $\sigma_S^2$ is 0.04 in Equation (19). A direct consequence of the assumption of a constant $\sigma_S^2$ is that the relationship between <Φ> and <*S*> is approximated with a single-valued function. In terms of evapotranspiration, the statistical dynamic approach better approximates the data from the mosaic approach for <*S*> values around <*S**> relative to the effective-parameter approach. However, this approach does not improve predictions of the leakage component of <Φ>.

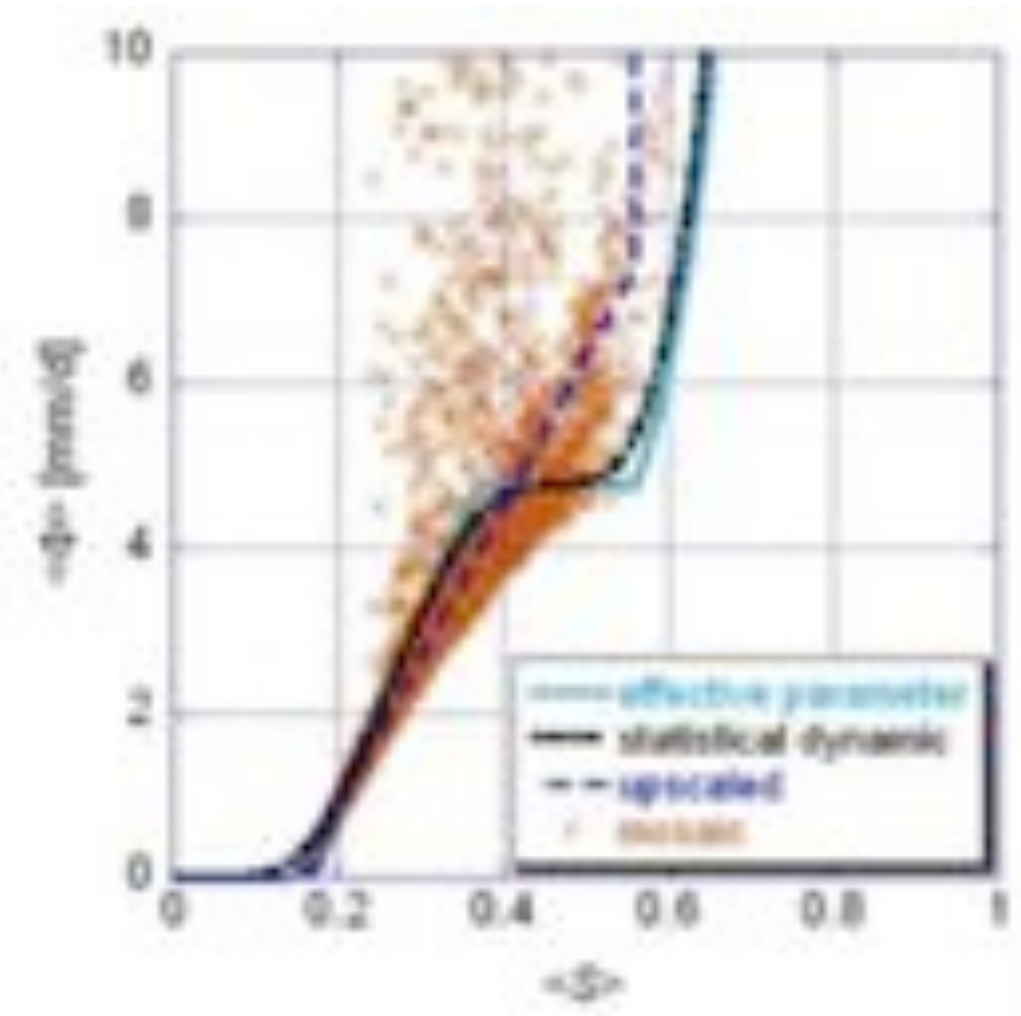

**Figure 10**: Comparison of the spatially averaged soil-water loss, <Φ>, versus <*S*> from the statistical dynamical approach with the relationships predicted by the effective parameter, upscaled, and mosaic approaches for a hypothetical Texas grassland assuming $\sigma_S^2$ is 0.04.

### *5.2. Montaldo & Albertson approach*

Montaldo and Albertson [33] presented a framework that predicts the temporal dynamics of the spatial variability of soil moisture and spatially averaged land-surface fluxes for ecosystems with vegetation and soil heterogeneity. In particular, spatially averaged land-surface fluxes (<*ET*>, <*L*>, etc.) are approximated through spatial averaging of 2nd order Taylor series expansions. These averaged land-surface fluxes are used in a water-balance equation to calculate <*S*>. Furthermore, the temporal evolution of the spatial variance of root-zone soil moisture is represented by a conservation equation derived by Albertson and Montaldo [1] as

$$Z_r \frac{\partial \sigma_S^2}{\partial t} = 2\left\langle S' \cdot q_{\text{in}}' \right\rangle - 2\left\langle S' \cdot ET' \right\rangle - 2\left\langle S' \cdot L' \right\rangle \tag{20}$$

where $q_{\text{in}}$ is rainfall minus interception ($R(t)$-$I[R(t)]$) and primed variables stand for plant-scale deviations from their spatial averages. The flux terms in Equation (20) depend on $S$ and plant-scale parameters characterizing the soil and vegetation (e.g. $ET_{\max}$, $S^*$, $S_w$, and $K_{sat}$), which we denote as $\vec{p} = [p, \ldots, p_N]$. Montaldo and Albertson [1] derived an expression for these plant-scale fluxes, $f(S, \vec{p})$, also using spatial averaging of a 2nd order Taylor series expansion. The expression is given by

$$\left\langle S' \cdot f_i' \right\rangle = \left.\frac{\partial f_i}{\partial S}\right|_{\langle S\rangle,\langle \vec{p}\rangle} \sigma_S^2 + \sum_{j=1}^{N} \left.\frac{\partial f_i}{\partial p_j}\right|_{\langle S\rangle,\langle \vec{p}\rangle} \left\langle S' \cdot p_j' \right\rangle + \frac{1}{2} \left.\frac{\partial f_i^2}{\partial S^2}\right|_{\langle S\rangle,\langle \vec{p}\rangle} \left\langle S'^3 \right\rangle$$
$$+ \frac{1}{2} \sum_{j=1}^{N} \sum_{k=1}^{N} \left.\frac{\partial^2 f_i}{\partial p_j \partial p_k}\right|_{\langle S\rangle,\langle \vec{p}\rangle} \left\langle S' \cdot p_j' \cdot p_k' \right\rangle + \sum_{j=1}^{N} \left.\frac{\partial^2 f_i}{\partial S \partial p_j}\right|_{\langle S\rangle,\langle \vec{p}\rangle} \left\langle S'^2 \cdot p_j' \right\rangle \tag{21}$$

It is important to note that this equation cannot be applied to the piecewise linear *ET* function in Equation (10), since Equation (21) is based on a Taylor series approximation.

The covariance terms ($\langle S' \cdot p'_j \rangle, \langle S' \cdot p'_j \cdot p'_k \rangle, \langle S'^2 \cdot p'_j \rangle$) and the third central moment of soil moisture, $\langle S'^3 \rangle$, in Equation (21) are a priori unknown. Montaldo and Albertson [1] found that the evolution of these terms through time is related to the dynamics of <*S*>. Based on the relationship between $\sigma_S^2$ and <*S*> in Fig. 9, hysteretic terms must arise in the equation for the temporal evolution of $\sigma_S^2$. In fact, Montaldo and Albertson [1] demonstrated through distributed model simulations (analogous to our mosaic approach) that non-unique, hysteretic relationships exist between these terms and <*S*>. However, the Monaldo & Albertson approach uses fitted polynomials rather than a hysteretic model to approximate the non-unique relationships. This approximation was justified based on evaluation of spatially averaged flux and $\sigma_S^2$ predictions.

### *5.3. Relevance of upscaled-parameter approach*

All of the approaches discussed in the paper approximate non-unique relationships that arise due to spatial averaging in the presence of heterogeneity with functions that do not model hysteresis. In particular, the upscale-parameter approach, presented herein, approximates the non-unique relationships between land-surface fluxes and <*S*>. For the Montaldo & Albertson approach, variables in the Taylor series expansions that have a non-unique dependence on <*S*> are approximated with polynomials. With regard to the statistical dynamic approach, the shape of pdfs of variables, affected by heterogeneity, do not evolve in time. Therefore, the non-unique relationships between shape parameters of a pdf and <*S*> are assumed equal to constant values.

The upscaled-parameter approach for dealing with spatial heterogeneity can be applied to any grid block in a large-scale model. The parameters of the upscaled functions (Table 4) can be categorized based on analysis of the relevant spatial scales of heterogeneity for a broad range of land-surface types and conditions, so that different functional relationships and parameters are not required for every grid block in a large-scale model. For each grid block, the explicit model must incorporate processes that are important to predict accurately the variables of interest, which are <*S*> and <*ET*> for this investigation. For example, if soil-moisture dynamics within a particular grid block are significantly affected by the topography of the land surface, then the mosaic approach must represent the effects of topography on soil moisture. These procedures are also necessary for the statistical dynamic and Montaldo & Albertson approach, since their parameters are also ecosystem dependent.

An important objective of this paper is to highlight the importance of relevant scales of heterogeneity within a grid block of a model for all of the approaches discussed. The vegetation heterogeneity occurs on scales much smaller than both the area of the modeled grid block and the scale of rainfall heterogeneity for the modeled ecosystem. Therefore, the upscaled relationships of the grid block will not be very sensitive to the spatial patterns of vegetation. Conversely, spatial structure of rainfall heterogeneity will have a significant effect on the upscaled relationships, because its spatial scale is of the same order of magnitude as the grid-block scale. Through identification of the relevant heterogeneity scales, we can anticipate the general characteristics and importance of non-unique relationships between spatially averaged quantities.

## 6. Conclusions and Future Research

Plant-scale relationships controlling the soil-water balance are generally not valid at larger spatial scales when spatial heterogeneity in rainfall and vegetation type exists. The relationships between spatially averaged variables controlling soil-moisture dynamics are non-unique, despite having no hysteresis at the plant scale. The characteristics of these non-unique relationships depend on the size of the averaging area and the spatial properties of the soil, vegetation, and rainfall. Plant-scale relationships can be upscaled to the scale of a regional land-surface model based on simulation data obtained through explicit representation of spatial heterogeneity in rainfall and plant type.

Excluding periods of immediately after large rainfall events, both approaches are relatively accurate in terms of the timing and magnitude of spatially averaged soil moisture and evapotranspiration predictions. The proposed upscaled function will improve predictions of these spatially averaged variables relative to the effective-parameter approach after large rainfall events, depending on the scales of heterogeneity and the size of the averaging area. The timing and magnitude of spatially averaged leakage predictions at the daily timescale are not significantly improved, although long-term predictions of spatially averaged leakage are more accurate.

The upscaled-parameter, statistical dynamic, and Montaldo & Albertson approaches use single-valued functions to approximate non-unique relationships between spatially averaged variables. These approaches must have the parameters of their functions estimated based on ecosystem properties, since the characteristics of the non-unique relationships change depending on the spatial scales of heterogeneity and the modeled grid block. Future work will focus on will focus on development of a multi-valued upscaled model between spatially averaged leakage and spatially averaged soil moisture in the spirit of Guswa [18].

## 7. Appendix A

Following the approach of Laio et al. [28], we present the analytical solution to Equation (13), which describes spatially averaged soil-moisture decay from an initial condition $\langle S_0 \rangle$ in the absence of rainfall:

$$\langle S(t)\rangle = \begin{cases} \left(\langle S_0\rangle + \frac{c_4}{c_3}\right)\exp\left(-\frac{c_3}{\langle \phi Z_r\rangle}\left(t - t_{\langle S_{fc}\rangle}\right)\right) - \frac{c_4}{c_3}, \\ \quad 0 \le t < t_{\langle S_{fc}\rangle}, \\ \left(\langle S_{fc}\rangle + \frac{c_2}{c_1}\right)\exp\left(-\frac{c_1}{\langle \phi Z_r\rangle}\left(t - t_{\langle S_w\rangle}\right)\right) - \frac{c_2}{c_1}, \\ \quad t_{\langle S_{fc}\rangle} \le t < t_{\langle S_w\rangle}, \\ \langle S_h\rangle + \left(\langle S_w\rangle - \langle S_h\rangle\right)\exp\left(-\frac{\langle E_w\rangle}{\langle \phi Z_r\rangle\left(\langle S_w\rangle - \langle S_h\rangle\right)}\left(t - t_{\langle S_w\rangle}\right)\right), \\ \quad t_{\langle S_w\rangle} \le t < \infty, \end{cases} \tag{A1}$$

where

$$t_{\langle S_{fc}\rangle} = \frac{-\langle \phi Z_r\rangle}{c_3}\ln\left(\frac{\langle S_{fc}\rangle + \frac{c_4}{c_3}}{\langle S_0\rangle + \frac{c_4}{c_3}}\right) \tag{A2}$$

$$t_{\langle S_w\rangle} = \frac{-\langle \phi Z_r\rangle}{c_1}\ln\left(\frac{\langle S_w\rangle + \frac{c_2}{c_1}}{\langle S_{fc}\rangle + \frac{c_2}{c_1}}\right) + t_{\langle S_{fc}\rangle} \tag{A3}$$

represent the time to evolve to $<S_{fc}>$ and $<S_w>$, respectively.

***Acknowledgements:***
The authors gratefully acknowledge the support of the Princeton Environmental Institute (PEI) fellowship.

**Table 1**: Parameters values for the woody plant *Prosopis glandulosa* and the $C_4$ grass *Paspalum setaceum* in sandy loam soil from Laio et al. [27], which were based on data from Hass & Dodd [21], Cuomo et al. [11], Ludlow [31], Stroh et al. [52], and Wan & Sosebee [55].

| | *Prosopis glandulosa* | *Paspalum setaceum* | Effective parameters for uniform rainfall | Effective parameter for non-uniform rainfall case (57% |
|---|---|---|---|---|
| $S_w$ | 0.180 | 0.167 | 0.176 | 0.176 |
| $S^*$ | 0.350 | 0.370 | 0.357 | 0.357 |
| $Z_r$ [mm] | 1000 | 400 | 700 | 660 |
| $ET_{max}$ | 4.42 | 4.76 | 4.59 | 4.61 |
| $E_w$ [mm/d] | 0.20 | 0.13 | 0.165 | 0.160 |
| $\Delta$ [mm] | 2.0 | 1.0 | 1.5 | 1.43 |

**Table 2**: Parameter values characterizing the temporal and spatial characteristics of thunderstorms during the growing season in eastern Rio Grande Plains of Texas.

| Parameter | Value |
|---|---|
| Storm arrival rate, $\lambda_t$ | 0.167 $d^{-1}$ |
| Mean rainfall depth at cell center, | 25.2 mm |
| Number of cells per area, $\lambda_{xy}$ | 0.0155 |
| Mean rainfall depth in subgrid | 15 mm |
| Storm cell parameter, $a'$ | 5 km |

**Table 3**: Soil parameters values for sandy loam in Texas from Laio et al. [27], which were based on data from US DOA [54] and Clapp & Hornberger [9].

| Parameter | Sandy Loam |
|---|---|
| Saturated conductivity, $K_{sat}$ | 822 mm/d |
| Porosity, $\phi$ | 0.43 |
| Soil retention parameter, $b$ | 4.9 |
| Relative soil-moisture at field capacity, $S_{fc}$ | 0.56 |
| Relative soil-moisture at hygroscopic point, $S_h$ | 0.14 |

**Table 4**: Parameters for the upscaled soil-water loss function in Eq. (14) and the upscaled <*ET*> function in Eq. (15).

| Parameter | Value |
|---|---|
| $c_1$ | 1830 mm/d |
| $c_2$ | -2.830 mm/d |
| $c_3$ | 1469 mm/d |
| $c_4$ | -781.7 mm/d |
| $c_5$ | 1.457 |
| $c_6$ | 2.689 |